# Direct Observation of a Semi-Bare Electron Coulomb Field Recover

**G Naumenko, Yu Popov and M Shevelev**
Tomsk Polytechnic University, Lenina str. 2, Tomsk, 634050, Russia

E-mail: naumenko@tpu.ru

**Abstract.** The problem of "semi-bare electron" was first considered in frame of quantum electrodynamics by E.L. Feinberg in 1980. In theory in frame of classical electrodynamics this problem was touched on in articles of N.F. Shul'ga and X. Artru. In 2008 the experimental investigations of this phenomenon in millimeter wavelength region were started by the group of scientists, including authors of this article. Used technique allowed us to study this effect in macroscopic mode. In this paper we present the results of direct observation of a semi-bare electron coulomb field recovery. The semi-bare state was obtained by passing of electron beam through the hole in a conductive screen. Measured spatial distribution of electromagnetic field shows the process of recover of the electron coulomb field, which is followed by a forward radiation. The experiments were performed on the relativistic electron beam of the microtron of Tomsk Polytechnic University.

## 1. About problem

The term "semi-bare electron" for the first time, probably was used in the article of E.L. Feinberg [1], where in the framework of quantum electrodynamics was considered the problem of scattering of a relativistic electron at large angle. The author has noted that the state just after the scattering may be interpreted as the state of the electron, which partially lost its own Coulomb field (the "half-naked electron" or "semi-bare electron"). In the framework of classical electrodynamics, this problem was raised in articles of N.F. Shulga [2] and X. Artru [3]. Most clearly, this problem may be represented in the view of the electron field as a field of pseudo-photons. The "pseudo-photon" method proposed by Fermi [4] and developed by Williams [5] is widely used for theoretical studies of electromagnetic processes (see for example [6] and [7]). According to this approach, the field of a charged particle may be replaced by a field of photons, which in this case are called pseudo-photons (in [8] is used the term "virtual quanta" (it should be differed on the same term in the quantum theory)). This approach provides a good accuracy for the ultra-relativistic particles when the particle velocity is close to the light velocity ($v \sim c$) (see [6]) and when the longitudinal electric field component of the particle is negligible. In this case the particle field has the same properties as the field of real photons.

Let us consider the interaction of real photons with a thick conductive mirror of high reflectivity. It is clear that real photons are reflected almost completely, they do not penetrate the target, and they do not excite surface currents on the downstream surface of the target. Because the properties of pseudo-photons are close to the properties of real photons, we may expect that with the passage of electrons through a hole in a conducting screen the part of the electron field is reflected from the screen and downstream to the screen the electron will lose a part of its Coulomb field. It is clear that in the further evolution the Coulomb field of the electron will be restored, because far from the screen we always observe electrons with the usual Coulomb field.

We take an interest in the study of this field and its spatial distribution. In [2] the solution of Maxwell's equations for a similar problem (the passage of electrons through the conducting screen) is presented. In this work, the electrons, emitted from a screen, are represented by the superposition of the theta function $\theta(t-t_0)$ and the expression for the current density of a uniformly moving electron, where $t_0$ is the emitting time of an electron from the screen. This statement, however, may not be used for our problem, because it is equivalent to consider a suddenly started electron, while we are considering the dynamics of the field of a uniformly moving electron.

The problem of field radiation of electrons emitted from the absorber in the form of "blackbody", was considered in [9], but the solution is done for the far zone (at infinity) and does not consider the dynamics of the field within the zone of radiation formation, which is more interesting for us.

In [10] we studied the dynamics of electrons in condition of a field of shadow effect from conducting and absorbing screen. The studies were based on the radiation characteristics of semi-bare electrons downstream to conductive target, or based on the evolution of the electron field pseudo-photons, diffracted on an electron dump, in which the electrons are stopped. These measurements are circumstantial and not give a full picture of the electron field distribution during the its evolution, although they give the general regularities in the field distribution.

In this article the direct measurement of the field characteristics of the electron bunch passing through a hole in conducting screen are presented. Such method will allow a direct study of the electron field during the transit from the state of semi-bare electron, to the usual (stable) state..

## 2. Experimental set-up and procedure

The experiment was carried out in the extracted electron beam of the Tomsk Nuclear Physics Institute microtron with parameters presented in Table 1.

**Table 1.** Electron beam parameters.

| Electron energy | 6.1 MeV ($\gamma=12$) | Bunch period | 380 psec |
|---|---|---|---|
| Train duration | $\tau \approx 4\ \mu\text{sec}$ | Bunch population | $N_e=6\cdot 10^8$ |
| Bunches in a train | $n_b \approx 1.6\cdot 10^4$ | Bunch length | $\sigma \approx 1.3\sim 1.6$mm |

For the electron field measurement we use the probe (figure 1) based on the well-known technique, which is applied for the surface current measurement in strip-line beam position monitors [6].

Beam parameters of microtron allow using coherent properties in interaction of electron field with probe in frequency region corresponding to the millimetre wavelength range. This possibility increases the probe response by the 8 orders (proportional to the bunch population) and makes this response achievable for measurements using existing sensors.

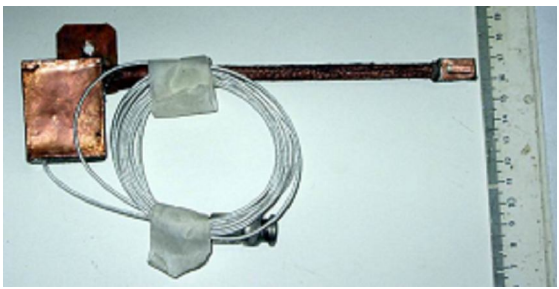

**Figure 1**. The view of probe for electron field measurement.

The probe has a maximum sensitivity for the Fourier harmonics of the field corresponding to a wavelength equal to 16 mm. The characteristics of the probe were tested on the stand (see figure 2) with a beam of real photons from the source with a wavelength equal to 11 mm.

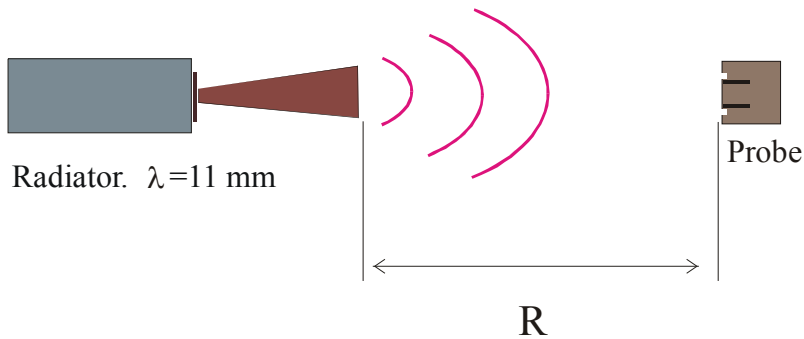

**Figure 2**. The scheme for testing of the probe properties.

On the stand was measured the probe response as a function of the distance R from the radiator (figure 3). To verify the nature of the probe response the values of the measured dependence was approximated by an exponential function $y = a \cdot x^{-b}$ (figure 3).

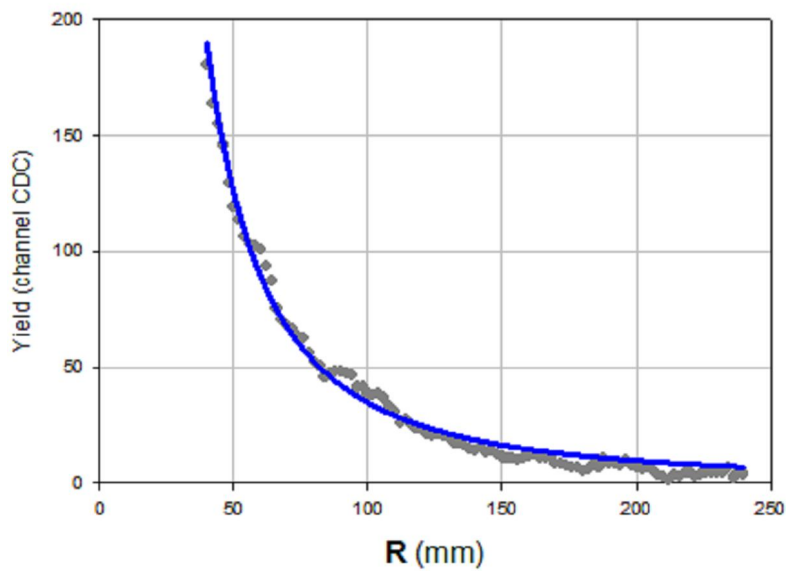

**Figure 3**. The probe response as a function of the distance R from the radiator. The points are the measured values and the sold line is the approximation $y = a \cdot x^{-b}$.

The value of exponential factor $b$ was estimated as $b = 1.86 \pm 0.05$. As is seen, the experimental dependence is well approximated by the inverse squared dependence. Up to a scale factor this corresponds to the measurement of squared field value.

### 3. Electron field distribution measurement

The measurements were performed using the scheme shown in figure 4. Relativistic electrons grouped in bunches of approximately $10^8$ electrons, passing through a hole in a conducting screen.

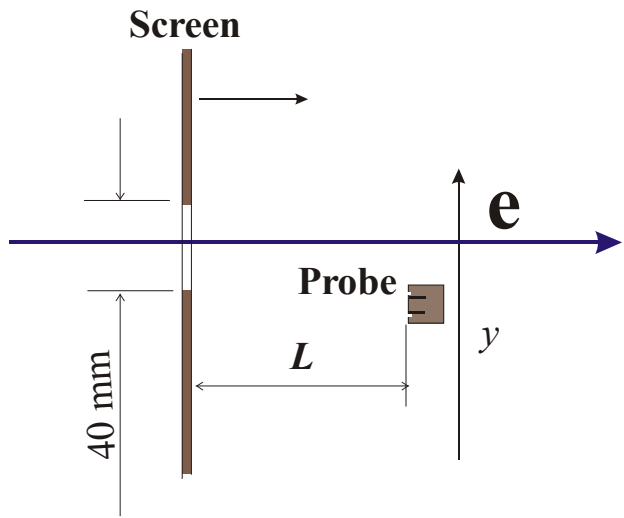

**Figure 4.** The scheme of electron field distribution measurement.

Most of the electron field pseudo-photons are reflected from the conducting screen. The remaining part of the Coulomb field of electrons passes through a hole in the screen. This forms an unstable state of the electrons, partially lost its own Coulomb field. As this state must be restored to a stable state of normal electrons during the further evolution of the electrons, the distribution of electrons field field must depend on the distance of the electron bunch from the screen. This dependence was measured in an experiment using a probe. Keep in mind that the spatial resolution of the probe is $\approx 4$ mm, so that within this region the probe detects the average value of the field. Then, due to the axial symmetry of the electron bunch field, the average field in the center of the bunch should be zero. Ie, the transverse field distribution of the electron bunch should have a dip in the center of the beam.

The field distribution was measured with step of 2 mm in the direction perpendicular to the beam and with step of 20 mm in variation of $L$. To exclude the beam divergence contribution, the probe was moved only in transversal direction to the electron beam. The distance $L$ was varied by moving the screen along the beam. The maximum value of $L$ was limited by a beam size in hole of screen due the beam divergence ($\approx 0.1$ radian in full). For measurement the background, the aluminium foil was glued on the active surface of the probe by the conductive adhesive. The statistical error of measurement was 5%. The image of the experimental set-up is shown in the figure 5.

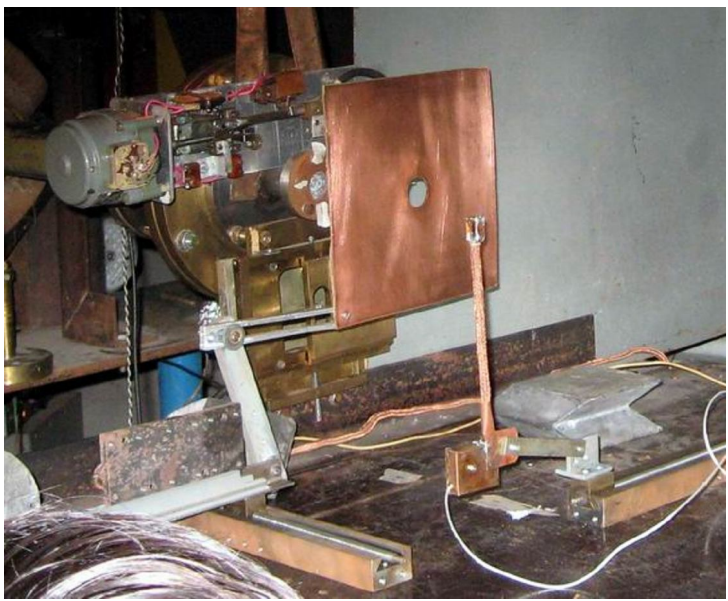

**Figure 5.** The view of the experimental set-up

The background was measured for the same area and was subtracted from the measured dependence. The smoothed dependence of the squared field on the transversal coordinate $y$ and on the distance $L$ between the screen and the probe is shown in figure 6.

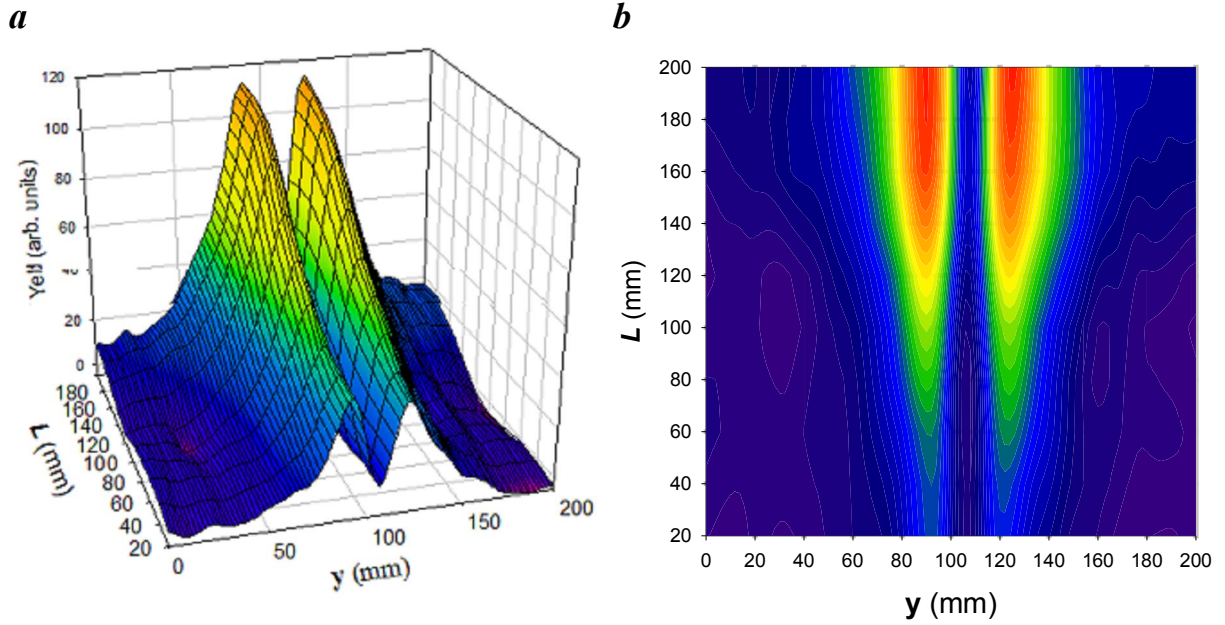

**Figure 6**. Smoothed experimental dependence of the squared field on the transversal coordinate $y$ and on the distance $L$ between the screen and the probe. $a$ – 3D view, $b$ - the plane representation of this dependence

**4. Discussion**
We see in figure 6 the build-up of the field with increasing of the distance $L$ from screen. One may say that we observe the destructive interference of the forward transition radiation from conductive screen with the field of electron (see [13]). However, according to our experimental result [12] the surface currents are not induced on a downstream surface of a thick conductive screen in investigated spectral region. Therefore no radiation may be emitted from the screen in this spectral region. It is clear from Fourier representation of Maxwell's equations together with continuously equation written as:

$$\begin{cases} \Delta \vec{E}_\omega + \omega^2 \vec{E}_\omega = 4\pi \left( \nabla \rho_\omega + i\omega \cdot \vec{j}_\omega \right) \\ \text{div}\, \vec{E}_\omega = 4\pi \rho_\omega \\ \text{div}\, \vec{j}_\omega + i\omega \cdot \rho_\omega = 0 \end{cases}$$

It is seen from this equation system, that if a surface current is absent in some frequency region then the system became homogeneous and describes only plane waves without radiation sources in this frequency region. This result is in agreement with pseudo-photon viewpoint in consideration of the electromagnetic field of relativistic electrons.

For analysis the processes in used geometry let us consider the ideal case when the relativistic electron moves along the axis $z$ through a thick conductive infinity screen which divides the space on the left and right half-spaces in plane $z = 0$. In this case we cannot use the model of suddenly accelerated electron, because in our experiment the electrons moves uniformly. If the electron moves in the left half-space, then no response may be observed in right half-space. Let the electron crosses the screen at the point of time $t = 0$. After this point the electron current appears in right half-space and no another currents are induced (see above). Therefore only electron current may be considered as a source of radiation in right half-space.
The electric field may be found from solution of Maxwell's equations in Fourier representation:

$$\vec{E} = \frac{i\omega}{k^2 - \omega^2}\left(\vec{j} - \frac{\vec{k}\cdot\vec{j}}{\omega^2}\vec{k}\right),$$

where $\vec{j}$ is the single electron current density.

For $0 < t < \infty$  $\vec{j}(\vec{k}) = \frac{e\vec{v}}{2\pi}\left(\frac{\delta(\omega + \vec{k}\cdot\vec{v})}{2} + i\frac{1}{2\pi(\omega + \vec{k}\cdot\vec{v})}\right),$ and for axial symmetry (as it is in our case) after inverse Fourier transform on z variable we obtain

$$E_\rho(z, k_\rho) = -i\frac{e}{8\pi^2}\frac{k_\rho}{k_\rho^2 + \frac{\omega^2}{\beta^2\gamma^2}}e^{i\frac{\omega z}{\beta}} + \frac{e\beta}{(2\pi)^3\omega}\int_{-\infty}^{\infty}\frac{e^{-ik_z z}}{k_z^2 + k_\rho^2 - \omega^2}\cdot\frac{k_z k_\rho}{\omega + k_z \cdot \beta}dk_z,$$

where $\rho$ is the transversal coordinate in axial system. Finally after full inverse transform we obtain

$$E_\rho(z, \rho) = \frac{1}{2\pi}\int_0^\infty E_\rho(z, k_\rho)\cdot J_0(k_\rho \cdot \rho)k_\rho dk_\rho, \qquad (1)$$

where $J_0(k_\rho \cdot \rho)$ is Bessel function.
The further analytical simplification of this expression is complicate and we present in figure 8*a* the numerical representation of $|E_\rho(z, \rho)|^2$ for the experimental conditions $\gamma = 12$ and $\lambda = 10 mm$ (the grid for $\rho = 0$ is skipped due to the singularity in the expression under the integral in this point).

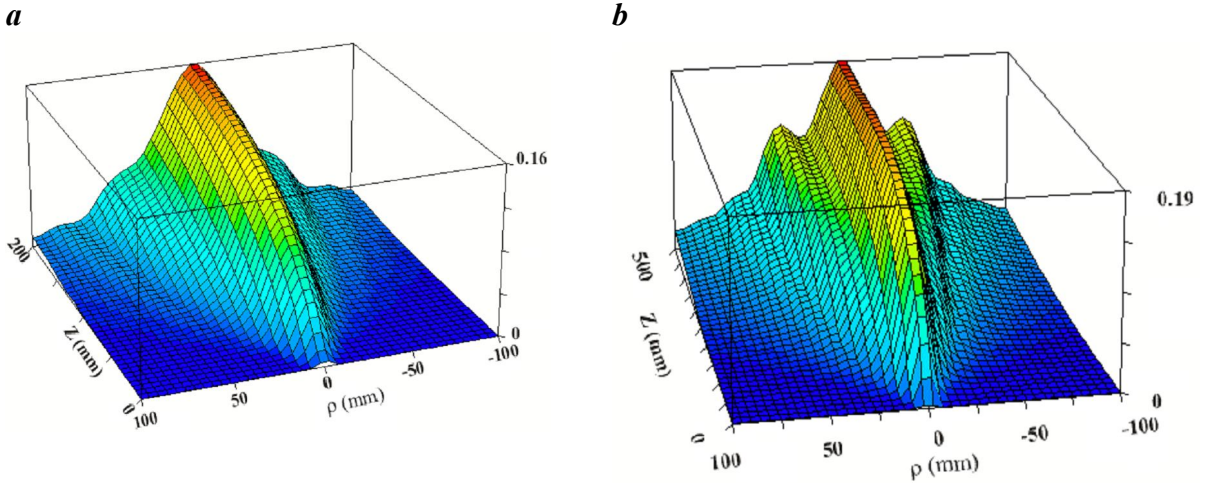

*a*  *b*

**Figure 8**. Numerical representation of $|E_\rho(z, \rho)|^2$ for experimental conditions.

We see in figure 8*a* that Maxwell's equation solution also shows the recovery of the field of electron crossing the conductive screen. For the larger distance from the screen (see figure 8*b*) we see the separation of the electron field and a radiation, which is named usually "transition radiation" and is usually supposed by mistake as the radiation from the screen.

In comparison with the experimental conditions, the theoretical calculations were performed for single electron (not for a bunch of finite size) crossed the screen with infinite small hole. Also the finite spatial resolution of the probe was not taken into ackount. Nevertheless both the experimental rasult and the theoretical calculation show the recovery of the electron field downstream to a conductive screen.

Finally we can conclude that we had observed in experiment the semi-bare electron and the process off electron field recovery in macroscopic mode.

**Acknowledgment**
This work was partly supported by the warrant-order 1.226.08 of the Ministry of Education and Science of the Russian Federation.